\RequirePackage{lineno}
\documentclass[twocolumn,preprintnumbers,amsmath,amssymb,floatfix]{revtex4}

\usepackage{graphicx}
\usepackage{dcolumn}
\usepackage{bm}
\usepackage{atlasphysics}
                         
\usepackage{amssymb}
\usepackage{subfigure}

\usepackage{hyperref}
                           
\begin{document}

\title{\vspace{-3.0cm}\flushleft{\normalsize\normalfont{CERN-PH-EP-2011-149}}
\flushright{\vspace{-0.86cm}\normalsize\normalfont{Submitted to Phys. Rev. Lett.}}
  \\ \center{Search for New Phenomena in \ttbar\ Events with Large
    Missing Transverse Momentum in Proton-Proton Collisions at
    $\sqrt{s}=7$ TeV with the ATLAS Detector}}

\author{ATLAS Collaboration}

\date{\today}

\begin{abstract}
  A search for new phenomena in \ttbar\ events with large missing
  transverse momentum in proton-proton collisions at a center-of-mass
  energy of 7 \TeV\ is presented. The measurement is based on
  1.04~\ifb\ of data collected with the ATLAS detector at the LHC.
  The search is carried out in the single-lepton channel,
  characterized by an isolated lepton of high transverse momentum,
  four or more jets and large missing transverse
  momentum. Contributions to this final state may arise from a number
  of Standard Model extensions. The results are interpreted in terms
  of a model where new top-quark partners are pair-produced and each
  decay to an on-shell top (or antitop) quark and a long-lived
  undetected neutral particle. The data are found to be consistent
  with Standard Model expectations. A limit at 95\% confidence level
  is set excluding a cross-section times branching ratio of 1.1~pb for
  a top-partner mass of 420~GeV and a neutral particle mass less than
  10~GeV. In a model of exotic fourth generation quarks, top-partner
  masses are excluded up to 420~GeV and neutral particle masses up to
  140~GeV.
\end{abstract}

\maketitle

The top quark holds great promise as a probe for new phenomena at the
TeV scale. It has the strongest coupling to the Standard Model Higgs
boson, and as a consequence it is the main contributor to the
quadratic divergence in the Higgs mass. Thus, assuming the
``naturalness'' hypothesis of effective quantum field theory, light
top partners (with masses below about 1 TeV) should correspond to one
of the most robust predictions of solutions to the hierarchy problem.

In this letter, a search is presented for pair-produced exotic top
partners $T\overline T$, each decaying to a top quark and a stable,
neutral weakly-interacting particle $A_0$, which in some models may be
its own antiparticle. The final state for such a process ($T\overline
T \to t \overline t A_0 A_0$) is identical to $t\overline{t}$, though
with a larger amount of missing transverse momentum (\met) from the
undetected $A_0$ pair. In supersymmetry models with $R$-parity
conservation, $T$ is identified with the stop squark and $A_0$ with
the lightest supersymmetric particle, the neutralino
($\chi_0$)~\cite{SUSYEllis} or the gravitino
($\tilde{G}$)~\cite{kats}. The $t\overline{t}+$\met~\cite{Han:2008gy}
signature appears in a general set of dark matter motivated models, as
well as in other Standard Model (SM) extensions, such as the
above-mentioned supersymmetry models, little Higgs models with
$T$-parity conservation~\cite{Cheng:2003ju,Cheng:2004yc,Cheng:2005as},
models of universal extra dimensions (UED) with Kaluza-Klein
parity~\cite{Appelquist:2000nn}, models in which baryon and lepton
number conservation arises from gauge
symmetries~\cite{FileviezPerez:2010gw} or models with third generation
scalar leptoquarks.  Many of these models provide a mechanism for
electroweak symmetry breaking and predict dark matter candidates,
which can be identified indirectly through their large \met\
signature.

The search is performed in the \ttbar\ single-lepton channel where one
$W$ boson produced by the top pair decays to a lepton-neutrino pair
($W\to \ell\nu$, including $\tau$ decays to $e$ or $\mu$) and the
other $W$ boson decays to a pair of quarks ($W\to q\overline q'$),
resulting in a final state with an isolated lepton of high transverse
momentum, four or more jets and large \met. The observed yield in this
signal region is compared with the SM expectation. In the absence of
signal an upper limit on the cross-section times branching ratio
BR($T\overline T \to t \overline t A_0 A_0$) is quoted. In the model
of exotic fourth generation up-type quarks~\cite{Alwall:2010jc} the
$T\overline T$ production cross-section is predicted to be
approximately six times higher than for stop squarks with a similar
mass~\cite{Han:2008gy}, due to the multiple spin states of two $T$'s
compared to scalar stops.  For this model the cross-section limits are
converted to an exclusion curve in the $T$ vs $A_0$ mass parameter
space.  A search for these exotic top-quark partners was performed in
proton-antiproton collisions at $\sqrt{s} = 1.96$~TeV by the CDF
Collaboration~\cite{Aaltonen:2011rr}.  The data were found to be
consistent with SM expectations. A 95\% confidence level limit was set
excluding a top-partner mass of 360~GeV for a neutral particle mass
less than 100~GeV. A recent update by CDF in the all-jets channel
excludes top-partner masses up to 400~GeV~\cite{Aaltonen:2011na}.

The ATLAS detector~\cite{AtlasDetector} consists of an inner detector
tracking system (ID) surrounded by a superconducting solenoid
providing a $2$~T magnetic field, electromagnetic and hadronic
calorimeters, and a muon spectrometer (MS).  The ID consists of pixel
and silicon microstrip detectors inside a transition radiation tracker
which provide tracking in the region $|\eta| < 2.5$~\cite{fn1}.  The
electromagnetic calorimeter is a lead/liquid-argon (LAr) detector in
the barrel ($|\eta| < 1.475$) and endcap ($1.375 < |\eta| < 3.2$)
regions.  Hadron calorimetry is based on two different detector
technologies. The barrel ($|\eta| < 0.8$) and extended barrel ($0.8 <
|\eta| < 1.7$) calorimeters are composed of scintillator/steel, while
the hadronic endcap calorimeters ($1.5 < |\eta| < 3.2$) are
copper/LAr.  The forward calorimeters ($3.1 < |\eta| < 4.9$) are
instrumented with copper/LAr and tungsten/LAr, providing
electromagnetic and hadronic energy measurements, respectively.  The
MS consists of three large superconducting toroids with 24 coils, a
system of trigger chambers, and precision tracking chambers which
provide muon momentum measurements up to $|\eta|$ of 2.7.

The analysis is based on data recorded by the ATLAS detector in 2011
using 1.04~\ifb\ of integrated luminosity.  The data were collected
using electron and muon triggers.  Requirements that ensure the
quality of beam conditions, detector performance and data are imposed.
Monte Carlo (MC) event samples with full ATLAS detector
simulation~\cite{AtlasSimulation} based on the {\sc Geant4}
program~\cite{Geant4} and corrected for all known detector effects are
used to model the signal process and most of the backgrounds. The
multi-jet background is modeled using data control samples rather than
the simulation.  The background sources are separated into four main
categories according to their importance: dilepton $t\overline{t}$
(where both $W$ bosons decay to a lepton-neutrino pair: $W\to
\ell\nu$); single-lepton $t\overline{t}$ and $W$+jets; multi-jet
production; and other electroweak processes, such as diboson
production, single top, and $Z$+jets.  The $t\overline{t}$ and single
top samples are produced with {\sc MC@NLO}~\cite{MC_at_NLO}, while the
$W$+jets and $Z$+jets samples are generated with {\sc
  Alpgen}~\cite{ALPGEN}. {\sc Herwig}~\cite{Herwig} is used to
simulate the parton shower and fragmentation, and {\sc
  Jimmy}~\cite{Jimmy} is used for the underlying event simulation. The
diboson background is simulated using {\sc Herwig}.  The
$t\overline{t}$ cross-section is normalized to approximate
next-to-next-to-leading order (NNLO) calculations~\cite{Moch:2008qy},
the inclusive $W$+jets and $Z$+jets cross-sections are normalized to
NNLO predictions~\cite{FEWZ}, and the cross-sections of the other
backgrounds are normalized to NLO predictions~\cite{MCFM}. Additional
corrections to the MC predictions are extracted from the data, as
described below.

Electron and muon candidates are selected as for other recent ATLAS
top quark studies using the single-lepton signature~\cite{Aad:2010ey}.
Jets are reconstructed using the anti-$k_t$~\cite{Cacciari:2008gp}
algorithm with the distance parameter $R = 0.4$.  To take into account
the differences in calorimeter response to electrons and hadrons, a
$\pt$- and $\eta$-dependent factor, derived from simulated events and
validated with data, is applied to each jet to provide an average
energy scale correction~\cite{JESv16} corresponding to the energies of
the reconstructed particles.

In the calorimeter, the energy deposited by particles is reconstructed
in three-dimensional clusters. These clusters are calibrated according
to the associated reconstructed high-\pt\ object.  The energy of these
clusters is summed vectorially, and projections of this sum in the
transverse plane correspond to the negative of the \met\
components~\cite{MET}.  Clusters not associated with any high-\pt\
object and muons reconstructed in the MS are also included in the
\met\ calculation.

Events are selected with exactly one isolated electron or muon that
passes the following kinematic selection criteria. Electrons are
required to satisfy $\et > 25$~GeV and $|\eta| < 2.47$. Electrons in
the region between the barrel and the endcap electromagnetic
calorimeters ($1.37 < |\eta| < 1.52$) are removed. Muon candidates are
required to satisfy $\pt > 20$~GeV and $|\eta| < 2.5$. These selected
leptons lie in the efficiency plateau of the single-lepton triggers.
Only events with four or more reconstructed jets with $\pt > 25$~GeV
and $|\eta| < 2.5$ are selected.  To reduce the single-lepton \ttbar\
and $W$+jets background, events are required to have $\met > 100$~GeV
and $m_{\mathrm{T}}$ $> 150$~GeV, where $m_{\mathrm{T}}$ is the
transverse mass of the lepton and \met~\cite{fn2}.  Events with either
a second lepton candidate with $\pt>15$~GeV or a track with
$\pt>12$~GeV, with no other tracks with $\pt>3$~GeV within $\Delta R =
0.4$ ($\Delta R\equiv \sqrt{\Delta\eta^2 + \Delta\phi^2}$), are
rejected in order to reduce the contribution from $t\overline{t}$
dilepton events. In particular the isolated track veto is useful in
reducing single-prong hadronic $\tau$ decays in $t\overline{t}$
dilepton events.  A summary of the background estimates and a
comparison with the observed number of selected events passing all
selection criteria are shown in Table~\ref{t:bg}.  A total yield of
101 $\pm$ 16 events is expected from SM sources, and 105 events are
observed in data.  The background composition is similar in the
electron and muon channels.

\begin{table}[ht!]
\begin{center}
  \caption[Background Summary]{Summary of expected SM yields including 
    statistical and systematic uncertainties compared with the 
    observed number of events in the signal region. 
  }
\begin{tabular}{r|c}
\hline\hline
Source & Number of events \\ 
\hline
Dilepton $t\overline{t}$ & 62 $\pm$ 15\\
Single-lepton $t\overline{t}$/$W$+jets & 33.1 $\pm$ 3.8 \\
Multi-jet & 1.2 $\pm$ 1.2\\
Single top & 3.5 $\pm$ 0.8\\
$Z$+jets & 0.9 $\pm$ 0.3\\
Dibosons & 0.9 $\pm$ 0.2\\
\hline
Total & 101 $\pm$ 16\\
\hline
Data  & 105\\
\hline\hline
\end{tabular}\label{t:bg}
\end{center}
\end{table}

The dominant background arises from $t\overline{t}$ dilepton final
states in which one of the leptons is not reconstructed, is outside
the detector acceptance, or is a $\tau$ lepton.  In all such cases,
the $t\overline{t}$ decay products include two high-\pt\ neutrinos,
resulting in large \met\ and $m_{\mathrm{T}}$ tails.  In MC, the
second lepton veto removes 45\% of the dilepton $t\overline{t}$ and
10\% of the single-lepton $t\overline{t}$ in the signal region.  The
veto performance is validated in the data in several control regions
both enhanced and depleted in dilepton $t\overline{t}$. Based on the
data-MC agreement in these control regions a 10\% uncertainty is
assigned to the veto efficiencies modeled in MC simulation.

The next largest background comes from single-lepton sources,
including $W$+jets and $t\overline{t}$ with one leptonic $W$ decay.
Both the normalization and the shape of the $m_{\mathrm{T}}$
distribution for this combined background are extracted from the data.
First, the yield of the single-lepton background estimated from
simulation is normalized in the control region 60~GeV $<$
$m_{\mathrm{T}}$ $<$ 90~GeV to the data which gives a correction of
($-5 \pm 3$)\%. Next, the shape of the $m_{\mathrm{T}}$ distribution
in MC is compared with data in various signal-depleted control
regions, where events satisfy the signal event selection but have
fewer than four jets.  In these control samples events with identified
$b$-jets, based on lifetime $b$-tagging~\cite{Aad:2010ey}, are
rejected in order to reduce the dilepton \ttbar\ background, such that
these control samples are dominated by $W$+jets events; the
corresponding loss of single-lepton \ttbar\ from this $b$-jet veto is
accounted for in the systematic uncertainties.  A comparison between
data and MC in this control region shows that MC systematically
underestimates the tails of the $m_{\mathrm{T}}$ distribution above
150 GeV, and a shape correction is derived that results in a ($15 \pm
10$)\% increase of the expected yield in the signal region.

The multi-jet background is extracted from the data using techniques
similar to those described in Ref.~\cite{Aad:2010ey}.  The techniques
exploit the fact that the lepton isolation efficiency is different in
signal and multi-jet events.  In both lepton channels the contribution
to the signal region is consistent with zero.

The contributions from single top, diboson production ($WW$, $WZ$, and
$ZZ$), and $Z$+jets are estimated using MC simulation, normalized to
the theoretical cross-section and total integrated luminosity.

The background yields estimated from MC simulated events are affected
by systematic uncertainties related to the modeling of detector
performance, reconstruction and object identification. The largest of
these uncertainties are from the jet energy scale~\cite{JESv16}
(approximately 5-7\% on the jet \pt, including a contribution from
pileup effects, leading to an 11\% uncertainty on the background event
yield), and from the performance of the second lepton veto in dilepton
$t\overline{t}$ (10\%).  Other uncertainties include those on the
lepton momentum scales and trigger and reconstruction
efficiencies. Lepton momentum scales and resolutions are determined
from fits to the $Z$-mass peak. Trigger and reconstruction
efficiencies are evaluated using tag-and-probe measurements in $Z
\rightarrow e^+e^-$\ or $Z \rightarrow \mu^+\mu^-$ events.  To
evaluate the effect of lepton momentum and jet energy scale
uncertainties, the \met\ and $m_{\mathrm{T}}$ are recalculated for
each uncertainty on selected objects.  Other small uncertainties
affecting the \met\ calculation are due to multiple $pp$ interactions,
jets with \pt\ below 20~\GeV, and calorimeter clusters that are not
associated to a selected object~\cite{METunc}.  Additionally,
theoretical cross-section uncertainties from choice of scales and
parton distribution functions are considered for these background
sources, as are the effects of using alternative MC generators, shower
models, and initial- and final-state radiation
tunings~\cite{Aad:2010ey}.  Finally, the 3.7\% uncertainty on the
integrated luminosity~\cite{bib:Lum2011} is applied to each background
source.

The systematic uncertainties applied to data-driven backgrounds are
determined from the data.  The dominant uncertainty for single-lepton
backgrounds is due to the ($15 \pm 10$)\% shape correction, and is
derived from the variation in the measured correction in different
control regions and from uncertainties in the $b$-tagging efficiency.
The uncertainty on the single-lepton normalization of ($-5 \pm 3$)\%
includes equal contributions from limited data statistics in the $W$
mass region and expected differences between the $W$+jets and
single-lepton $t\overline{t}$ contributions to the signal and control
regions.  A 100\% systematic uncertainty is assigned to the small
estimated multi-jet yield.

The expected and observed event yields are consistent within
statistical and systematic uncertainties.  Therefore, the results are
interpreted as a limit on the possible non-SM contribution to the
selected sample. A model involving pair-production of heavy quark-like
objects ($T\overline T$), each forced to decay to a top quark and a
scalar neutral $A_0$, is chosen to establish these limits.

{\sc MadGraph}~\cite{MadGraph} is used to simulate the signal process
with the parton distribution function set CTEQ6L1~\cite{CTEQ6L1}, and
{\sc Pythia}~\cite{Pythia} is used to simulate the parton shower and
fragmentation.  A grid of $T$ and $A_0$ masses is generated with
300~GeV $\le m(T) \le$ 450~GeV and 10~GeV $\le m(A_0) \le$ 150~GeV.
Each sample is normalized to the cross-section calculated at
approximate NNLO in QCD using {\sc Hathor}~\cite{Hathor}, ranging from
8.0 pb for a $T$ mass of 300~GeV to 0.66 pb for a $T$ mass of 450~GeV.
Using this grid of signal samples, the efficiency times acceptance for
the $T\overline{T}$ signal model is parametrized as a function of the
$T$ and $A_0$ masses to generate the expected signal event yield for
any pair of masses.  The combined acceptance times signal selection
efficiency varies between 3 and 5\% for small $A_0$ masses and
decreases to between 2 and 4\% for larger $A_0$ masses.

All common systematic uncertainties for MC-based backgrounds are
applied to this signal model. These include the uncertainties on the
jet energy scale, lepton reconstruction efficiencies and scales,
integrated luminosity and the dilepton veto efficiency. Overall, the
systematic uncertainty on the signal acceptance times efficiency
varies between 11 and 14\%, and is largest for those samples with a
$T$-$A_0$ mass difference closest to the top quark mass. The
theoretical uncertainties on the signal cross-section vary between 10
to 15\% and originate mainly from the choice of scales ($m_T/2 <
\mu_R=\mu_F < 2m_T$, where $\mu_R$ and $\mu_F$ are the renormalisation
and factorization scale) and parton distribution functions.

The \met\ and $m_{\mathrm{T}}$ distributions for data are shown in
Fig.~\ref{f:dataplots} and compared with the background and signal
predictions. There is no significant evidence of an excess over the SM
prediction, and the kinematics are well modeled.

\begin{figure}[h!]
  \centering
    \includegraphics[width=1.0\columnwidth]{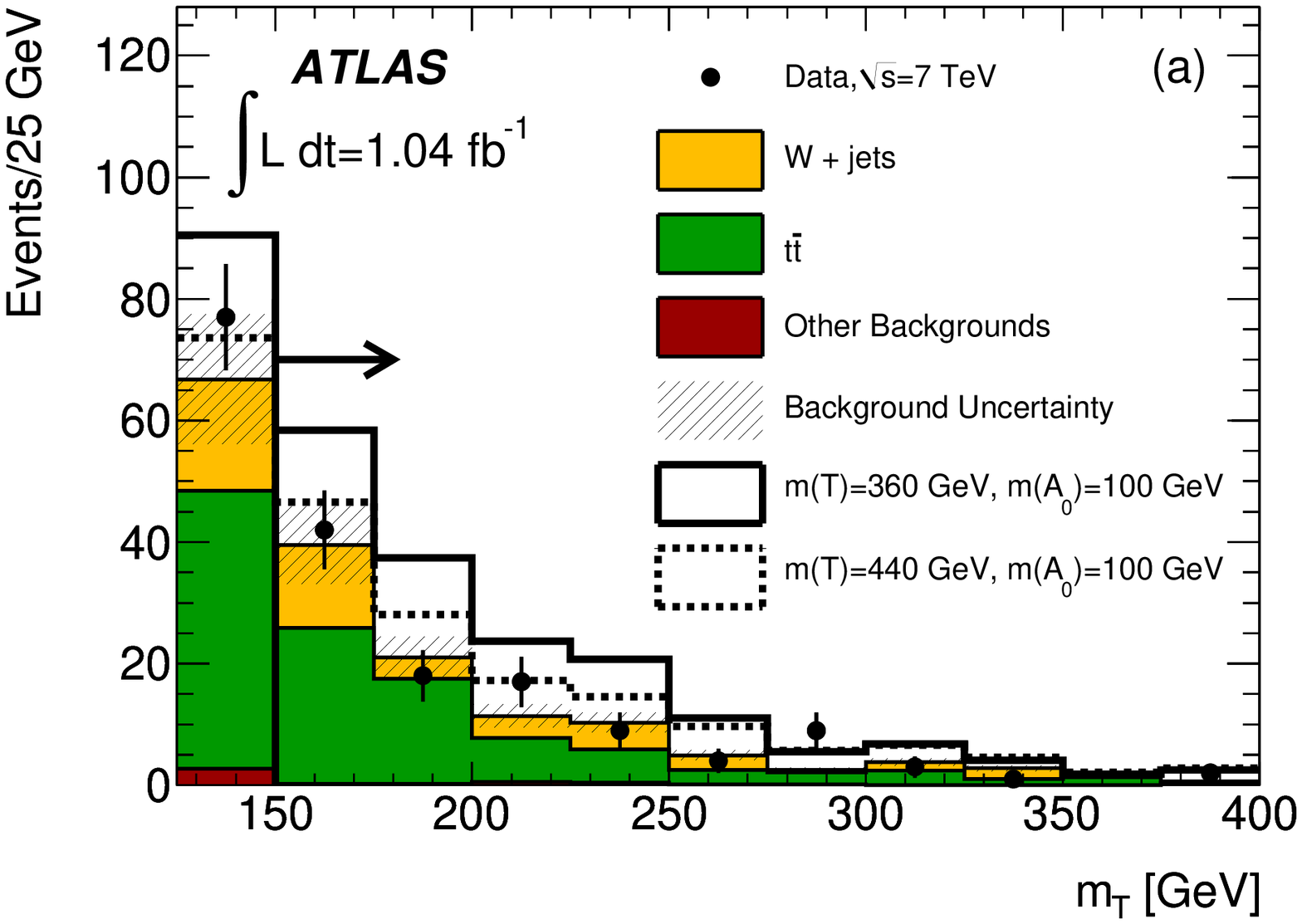}
    \includegraphics[width=1.0\columnwidth]{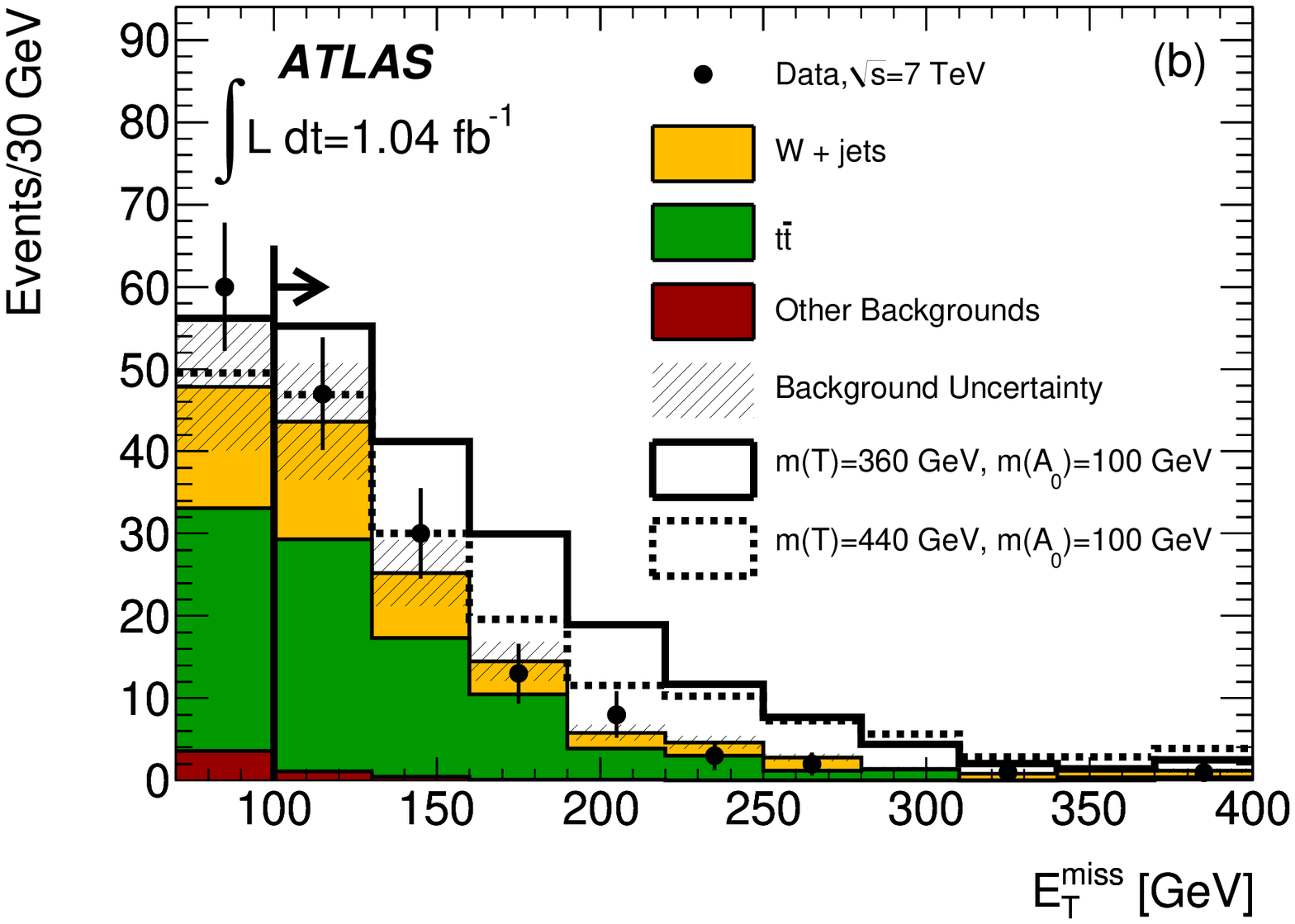}
    \caption{(a) Transverse mass of the lepton and missing energy and
      (b) \met\ after applying all selection criteria except the cut
      on the variable shown.  MC background contributions are stacked
      on top of each other and normalized according to the data-driven
      corrections discussed in the text.  The lines with the arrows
      indicate the selection criteria that define the signal region
      ($m_{\mathrm{T}}>150$~GeV and $\met>100$~GeV). `Other
      Backgrounds' includes both multi-jet backgrounds and $Z$+jets,
      single top and diboson production.  Expectations from two signal
      mass points are stacked separately on top of the SM background.
      The last bin includes the overflow.}
\label{f:dataplots}
\end{figure}
From the observed event yield and the predicted signal and background
event yields after all cuts, a frequentist confidence interval on the
signal hypothesis is calculated for various assumed $T$ and $A_0$
masses, assuming Gaussian systematic uncertainties.  Correlations
between signal and background uncertainties are included.
Figure~\ref{f:excl} shows the region of parameter space excluded at
the 95\% confidence level.  As the mass difference between the $T$ and
$A_0$ approaches the top quark mass, the $A_0$ contributes less
momentum to the \met, and signal becomes indistinguishable from SM
$t\overline{t}$. Assuming a $T\overline T \to t \overline t A_0 A_0$
branching ratio of 100\%, signal points with $T$ mass up to 420~GeV
are excluded at the 95\% confidence level for an $A_0$ mass below
10~GeV, as are signal points with 330~GeV~$< m(T) < 390$~GeV for an
$A_0$ mass below 140~GeV.  Figure~\ref{f:exclXS} shows the
cross-section times branching ratio excluded at the 95\% confidence
level versus $T$ mass, for an $A_0$ mass of 10 GeV.  A cross-section
times branching ratio of 1.1 (1.9)~pb is excluded at the 95\%
confidence level for a $T$ mass of 420 (370)~GeV and an $A_0$ mass of
10 (140)~GeV.  The estimated acceptance times efficiency for
spin-$\frac{1}{2}$ $T\overline{T}$ models is consistent within
systematic uncertainties with that for scalar models, such as pair
production of stop squarks (with a $t \overline t \chi_0 \chi_0$ final
state) or third-generation leptoquarks (with a $t \overline t
\nu_{\tau} \overline \nu_{\tau}$ final state). The cross-section
limits presented in Fig.~\ref{f:exclXS} are therefore approximately
valid for such models, although the predicted cross-section is
typically below the current sensitivity.

\begin{figure}[h!]
  \centering
  \includegraphics[width=1.0\columnwidth]{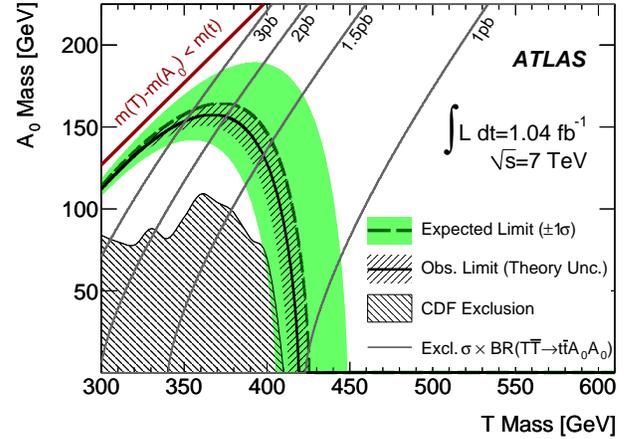}
  \caption{Excluded region (under the curve) at the 95\% confidence
    level as a function of $T$ and $A_0$ masses, compared with the CDF
    exclusion~\cite{Aaltonen:2011rr,Aaltonen:2011na}. Theoretical
    uncertainties on the $T\overline{T}$ cross-section are not
    included in the limit, but the effect of these uncertainties is
    shown.  The gray contours show the excluded cross-section times
    branching ratio as a function of the two masses. }
\label{f:excl}
\end{figure}

\begin{figure}[h!]
  \centering
    \includegraphics[width=1.0\columnwidth]{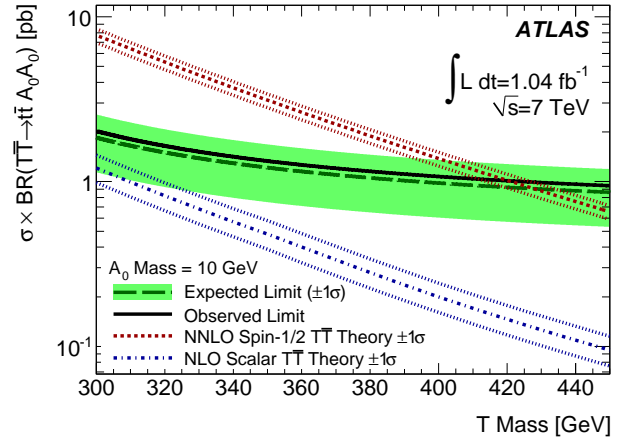}
    \caption{Cross-section times branching ratio excluded at the 95\%
      confidence level versus $T$ mass for an $A_0$ mass of 10~GeV.
      Theoretical predictions for both spin-$\frac{1}{2}$ and scalar
      $T$ pair production are also shown.}
\label{f:exclXS}
\end{figure}

In summary, in 1.04 fb$^{-1}$ of data in $pp$ collisions at a
center-of-mass energy of 7 TeV, there is no evidence of an excess of
events with large \met\ in a sample dominated by $t\overline{t}$
events. Using a model of pair-produced quark-like objects decaying to
a top quark and a heavy neutral particle, a limit is established
excluding masses of these top partners up to 420~GeV and stable
weakly-interacting particle masses up to 140~GeV (see
Fig.~\ref{f:excl}).  In particular, a cross-section times branching
ratio of 1.1~pb is excluded at the 95\% confidence level for
$m(T)=$~420~GeV and $m(A_0)=$~10~GeV.  The cross-section limits are
approximately valid for a number of models of new phenomena.

We thank CERN for the very successful operation of the LHC, as well as
the support staff from our institutions without whom ATLAS could not
be operated efficiently.  We acknowledge the support of ANPCyT,
Argentina; YerPhI, Armenia; ARC, Australia; BMWF, Austria; ANAS,
Azerbaijan; SSTC, Belarus; CNPq and FAPESP, Brazil; NSERC, NRC and
CFI, Canada; CERN; CONICYT, Chile; CAS, MOST and NSFC, China;
COLCIENCIAS, Colombia; MSMT CR, MPO CR and VSC CR, Czech Republic;
DNRF, DNSRC and Lundbeck Foundation, Denmark; ARTEMIS, European Union;
IN2P3-CNRS, CEA-DSM/IRFU, France; GNAS, Georgia; BMBF, DFG, HGF, MPG
and AvH Foundation, Germany; GSRT, Greece; ISF, MINERVA, GIF, DIP and
Benoziyo Center, Israel; INFN, Italy; MEXT and JSPS, Japan; CNRST,
Morocco; FOM and NWO, Netherlands; RCN, Norway; MNiSW, Poland; GRICES
and FCT, Portugal; MERYS (MECTS), Romania; MES of Russia and ROSATOM,
Russian Federation; JINR; MSTD, Serbia; MSSR, Slovakia; ARRS and MVZT,
Slovenia; DST/NRF, South Africa; MICINN, Spain; SRC and Wallenberg
Foundation, Sweden; SER, SNSF and Cantons of Bern and Geneva,
Switzerland; NSC, Taiwan; TAEK, Turkey; STFC, the Royal Society and
Leverhulme Trust, United Kingdom; DOE and NSF, United States of
America.  The crucial computing support from all WLCG partners is
acknowledged gratefully, in particular from CERN and the ATLAS Tier-1
facilities at TRIUMF (Canada), NDGF (Denmark, Norway, Sweden),
CC-IN2P3 (France), KIT/GridKA (Germany), INFN-CNAF (Italy), NL-T1
(Netherlands), PIC (Spain), ASGC (Taiwan), RAL (UK) and BNL (USA) and
in the Tier-2 facilities worldwide.

\bibliography{tt_met_pub}

\begin{thebibliography}{33}
\expandafter\ifx\csname natexlab\endcsname\relax\def\natexlab#1{#1}\fi
\expandafter\ifx\csname bibnamefont\endcsname\relax
  \def\bibnamefont#1{#1}\fi
\expandafter\ifx\csname bibfnamefont\endcsname\relax
  \def\bibfnamefont#1{#1}\fi
\expandafter\ifx\csname citenamefont\endcsname\relax
  \def\citenamefont#1{#1}\fi
\expandafter\ifx\csname url\endcsname\relax
  \def\url#1{\texttt{#1}}\fi
\expandafter\ifx\csname urlprefix\endcsname\relax\def\urlprefix{URL }\fi
\providecommand{\bibinfo}[2]{#2}
\providecommand{\eprint}[2][]{\url{#2}}

\bibitem[{\citenamefont{{J. Ellis, K. A. Olive}}(2010)}]{SUSYEllis}
\bibinfo{author}{\bibnamefont{{J. Ellis, K. A. Olive}}} (\bibinfo{year}{2010}),
  \eprint{arXiv:1001.3651}.

\bibitem[{\citenamefont{{Y. Kats, D. Shih}}(2011)}]{kats}
\bibinfo{author}{\bibnamefont{{Y. Kats, D. Shih}}}, \bibinfo{journal}{J. High
  Energy Phys.} \textbf{\bibinfo{volume}{1108}}, \bibinfo{pages}{049}
  (\bibinfo{year}{2011}).

\bibitem[{\citenamefont{{T.~Han, R.~Mahbubani, D.~G.~E.~Walker and
  L.~T.~E.~Wang}}(2009)}]{Han:2008gy}
\bibinfo{author}{\bibnamefont{{T.~Han, R.~Mahbubani, D.~G.~E.~Walker and
  L.~T.~E.~Wang}}}, \bibinfo{journal}{J. High Energy Phys.}
  \textbf{\bibinfo{volume}{0905}}, \bibinfo{pages}{117} (\bibinfo{year}{2009}).

\bibitem[{\citenamefont{{H.~C.~Cheng and I.~Low}}(2003)}]{Cheng:2003ju}
\bibinfo{author}{\bibnamefont{{H.~C.~Cheng and I.~Low}}}, \bibinfo{journal}{J.
  High Energy Phys.} \textbf{\bibinfo{volume}{0309}}, \bibinfo{pages}{051}
  (\bibinfo{year}{2003}).

\bibitem[{\citenamefont{{H.~C.~Cheng and I.~Low}}(2004)}]{Cheng:2004yc}
\bibinfo{author}{\bibnamefont{{H.~C.~Cheng and I.~Low}}}, \bibinfo{journal}{J.
  High Energy Phys.} \textbf{\bibinfo{volume}{0408}}, \bibinfo{pages}{061}
  (\bibinfo{year}{2004}).

\bibitem[{\citenamefont{{H.~C.~Cheng, I.~Low and
  L.~T.~Wang}}(2006)}]{Cheng:2005as}
\bibinfo{author}{\bibnamefont{{H.~C.~Cheng, I.~Low and L.~T.~Wang}}},
  \bibinfo{journal}{Phys. Rev. D} \textbf{\bibinfo{volume}{74}},
  \bibinfo{pages}{055001} (\bibinfo{year}{2006}).

\bibitem[{\citenamefont{{T.~Appelquist, H.~C.~Cheng and
  B.~A.~Dobrescu}}(2001)}]{Appelquist:2000nn}
\bibinfo{author}{\bibnamefont{{T.~Appelquist, H.~C.~Cheng and
  B.~A.~Dobrescu}}}, \bibinfo{journal}{Phys. Rev. D}
  \textbf{\bibinfo{volume}{64}}, \bibinfo{pages}{035002}
  (\bibinfo{year}{2001}).

\bibitem[{\citenamefont{{P.~Fileviez Perez and
  M.~B.~Wise}}(2010)}]{FileviezPerez:2010gw}
\bibinfo{author}{\bibnamefont{{P.~Fileviez Perez and M.~B.~Wise}}},
  \bibinfo{journal}{Phys. Rev. D} \textbf{\bibinfo{volume}{82}},
  \bibinfo{pages}{011901} (\bibinfo{year}{2010}).

\bibitem[{\citenamefont{{J.~Alwall, J.~L.~Feng, J.~Kumar and
  S.~Su}}(2010)}]{Alwall:2010jc}
\bibinfo{author}{\bibnamefont{{J.~Alwall, J.~L.~Feng, J.~Kumar and S.~Su}}},
  \bibinfo{journal}{Phys. Rev. D} \textbf{\bibinfo{volume}{81}},
  \bibinfo{pages}{114027} (\bibinfo{year}{2010}).

\bibitem[{\citenamefont{{T.~Aaltonen {\it et al.}, CDF
  Collaboration}}(2011{\natexlab{a}})}]{Aaltonen:2011rr}
\bibinfo{author}{\bibnamefont{{T.~Aaltonen {\it et al.}, CDF Collaboration}}},
  \bibinfo{journal}{Phys. Rev. Lett.} \textbf{\bibinfo{volume}{106}},
  \bibinfo{pages}{191801} (\bibinfo{year}{2011}{\natexlab{a}}).

\bibitem[{\citenamefont{{T.~Aaltonen {\it et al.}, CDF
  Collaboration}}(2011{\natexlab{b}})}]{Aaltonen:2011na}
\bibinfo{author}{\bibnamefont{{T.~Aaltonen {\it et al.}, CDF Collaboration}}}
  (\bibinfo{year}{2011}{\natexlab{b}}), \eprint{arXiv:1107.3574}.

\bibitem[{\citenamefont{{ATLAS Collaboration}}(2008)}]{AtlasDetector}
\bibinfo{author}{\bibnamefont{{ATLAS Collaboration}}}, \bibinfo{journal}{J.
  Instrum.} \textbf{\bibinfo{volume}{3}}, \bibinfo{pages}{S08003}
  (\bibinfo{year}{2008}).

\bibitem[{fn1()}]{fn1}
\bibinfo{title}{The azimuthal angle $\phi$ is measured around the beam
  axis and the polar angle $\theta$ is the angle from the beam axis. the
  pseudorapidity is defined as $\eta \equiv -\ln \tan(\theta/2)$.}

\bibitem[{\citenamefont{{ATLAS
  Collaboration}}(2010{\natexlab{a}})}]{AtlasSimulation}
\bibinfo{author}{\bibnamefont{{ATLAS Collaboration}}}, \bibinfo{journal}{Eur.
  Phys. J. C} \textbf{\bibinfo{volume}{70}}, \bibinfo{pages}{823}
  (\bibinfo{year}{2010}{\natexlab{a}}).

\bibitem[{\citenamefont{Agostinelli et~al.}(2003)}]{Geant4}
\bibinfo{author}{\bibfnamefont{S.}~\bibnamefont{Agostinelli}}
  \bibnamefont{et~al.}, \bibinfo{journal}{Nucl. Instrum. Methods Phys. Res.,
  Sect. A} \textbf{\bibinfo{volume}{506}}, \bibinfo{pages}{250}
  (\bibinfo{year}{2003}).

\bibitem[{\citenamefont{Frixione and Webber}(2002)}]{MC_at_NLO}
\bibinfo{author}{\bibfnamefont{S.}~\bibnamefont{Frixione}} \bibnamefont{and}
  \bibinfo{author}{\bibfnamefont{B.~R.} \bibnamefont{Webber}},
  \bibinfo{journal}{J. High Energy Phys.} \textbf{\bibinfo{volume}{0206}},
  \bibinfo{pages}{029} (\bibinfo{year}{2002}).

\bibitem[{\citenamefont{Mangano et~al.}(2003)}]{ALPGEN}
\bibinfo{author}{\bibfnamefont{M.~L.} \bibnamefont{Mangano}}
  \bibnamefont{et~al.}, \bibinfo{journal}{J. High Energy Phys.}
  \textbf{\bibinfo{volume}{0307}}, \bibinfo{pages}{001} (\bibinfo{year}{2003}).

\bibitem[{\citenamefont{Corcella et~al.}(2001)}]{Herwig}
\bibinfo{author}{\bibfnamefont{G.}~\bibnamefont{Corcella}}
  \bibnamefont{et~al.}, \bibinfo{journal}{J. High Energy Phys.}
  \textbf{\bibinfo{volume}{0101}}, \bibinfo{pages}{010} (\bibinfo{year}{2001}).

\bibitem[{\citenamefont{Butterworth et~al.}(1996)\citenamefont{Butterworth,
  Forshaw, and Seymour}}]{Jimmy}
\bibinfo{author}{\bibfnamefont{J.~M.} \bibnamefont{Butterworth}},
  \bibinfo{author}{\bibfnamefont{J.~R.} \bibnamefont{Forshaw}},
  \bibnamefont{and} \bibinfo{author}{\bibfnamefont{M.~H.}
  \bibnamefont{Seymour}}, \bibinfo{journal}{Z. Phys. C}
  \textbf{\bibinfo{volume}{72}}, \bibinfo{pages}{637} (\bibinfo{year}{1996}).

\bibitem[{\citenamefont{Moch and Uwer}(2008)}]{Moch:2008qy}
\bibinfo{author}{\bibfnamefont{S.}~\bibnamefont{Moch}} \bibnamefont{and}
  \bibinfo{author}{\bibfnamefont{P.}~\bibnamefont{Uwer}},
  \bibinfo{journal}{Phys. Rev. D} \textbf{\bibinfo{volume}{78}},
  \bibinfo{pages}{034003} (\bibinfo{year}{2008}).

\bibitem[{\citenamefont{Melnikov and Petriello}(2006)}]{FEWZ}
\bibinfo{author}{\bibfnamefont{K.}~\bibnamefont{Melnikov}} \bibnamefont{and}
  \bibinfo{author}{\bibfnamefont{F.}~\bibnamefont{Petriello}},
  \bibinfo{journal}{Phys. Rev. D} \textbf{\bibinfo{volume}{74}},
  \bibinfo{pages}{114017} (\bibinfo{year}{2006}).

\bibitem[{\citenamefont{Campbell et~al.}(2003)\citenamefont{Campbell, Ellis,
  and Rainwater}}]{MCFM}
\bibinfo{author}{\bibfnamefont{J.~M.} \bibnamefont{Campbell}},
  \bibinfo{author}{\bibfnamefont{R.~K.} \bibnamefont{Ellis}}, \bibnamefont{and}
  \bibinfo{author}{\bibfnamefont{D.~L.} \bibnamefont{Rainwater}},
  \bibinfo{journal}{Phys. Rev. D} \textbf{\bibinfo{volume}{68}},
  \bibinfo{pages}{094021} (\bibinfo{year}{2003}).

\bibitem[{\citenamefont{{ATLAS
  Collaboration}}(2011{\natexlab{a}})}]{Aad:2010ey}
\bibinfo{author}{\bibnamefont{{ATLAS Collaboration}}}, \bibinfo{journal}{Eur.
  Phys. J. C} \textbf{\bibinfo{volume}{71}}, \bibinfo{pages}{1577}
  (\bibinfo{year}{2011}{\natexlab{a}}).

\bibitem[{\citenamefont{Cacciari et~al.}(2008)\citenamefont{Cacciari, Salam,
  and Soyez}}]{Cacciari:2008gp}
\bibinfo{author}{\bibfnamefont{M.}~\bibnamefont{Cacciari}},
  \bibinfo{author}{\bibfnamefont{G.~P.} \bibnamefont{Salam}}, \bibnamefont{and}
  \bibinfo{author}{\bibfnamefont{G.}~\bibnamefont{Soyez}}, \bibinfo{journal}{J.
  High Energy Phys.} \textbf{\bibinfo{volume}{0804}}, \bibinfo{pages}{063}
  (\bibinfo{year}{2008}).

\bibitem[{\citenamefont{{ATLAS Collaboration}}(2011{\natexlab{b}})}]{JESv16}
\bibinfo{author}{\bibnamefont{{ATLAS Collaboration}}},
  \bibinfo{journal}{ATLAS-CONF-2011-032}  (\bibinfo{year}{2011}{\natexlab{b}}).

\bibitem[{\citenamefont{{ATLAS Collaboration}}(2010{\natexlab{b}})}]{MET}
\bibinfo{author}{\bibnamefont{{ATLAS Collaboration}}}, \bibinfo{journal}{J.
  High Energy Phys.} \textbf{\bibinfo{volume}{1012}}, \bibinfo{pages}{060}
  (\bibinfo{year}{2010}{\natexlab{b}}).

\bibitem[{fn2()}]{fn2}
\bibinfo{title}{The transverse mass is defined by the formula $
  m_{\mathrm{t}} = \sqrt{2p^{\ell}_T \met (1 - \cos(\phi^{\ell} - \phi^{\met}))
  }$, where $p^{\ell}_t$ is the \pt\ (\et) of the muon (electron) and
  $\phi^{\ell}$ ($\phi^{\met}$) is the azimuthal angle of the lepton (\met).}

\bibitem[{\citenamefont{{ATLAS Collaboration}}(2011{\natexlab{c}})}]{METunc}
\bibinfo{author}{\bibnamefont{{ATLAS Collaboration}}}
  (\bibinfo{year}{2011}{\natexlab{c}}), \bibinfo{note}{{ATLAS-CONF-2011-080}}.

\bibitem[{\citenamefont{{ATLAS
  Collaboration}}(2011{\natexlab{d}})}]{bib:Lum2011}
\bibinfo{author}{\bibnamefont{{ATLAS Collaboration}}}
  (\bibinfo{year}{2011}{\natexlab{d}}), \bibinfo{note}{{ATLAS-CONF-2011-116}}.

\bibitem[{\citenamefont{{J. Alwall {\it et al.}}}(2007)}]{MadGraph}
\bibinfo{author}{\bibnamefont{{J. Alwall {\it et al.}}}}, \bibinfo{journal}{J.
  High Energy Phys.} \textbf{\bibinfo{volume}{0709}}, \bibinfo{pages}{028}
  (\bibinfo{year}{2007}).

\bibitem[{\citenamefont{Pumplin et~al.}(2002)}]{CTEQ6L1}
\bibinfo{author}{\bibfnamefont{J.}~\bibnamefont{Pumplin}} \bibnamefont{et~al.},
  \bibinfo{journal}{J. High Energy Phys.} \textbf{\bibinfo{volume}{0207}},
  \bibinfo{pages}{012} (\bibinfo{year}{2002}).

\bibitem[{\citenamefont{Sj{\"{o}}strand
  et~al.}(2006)\citenamefont{Sj{\"{o}}strand, Mrenna, and Skands}}]{Pythia}
\bibinfo{author}{\bibfnamefont{T.}~\bibnamefont{Sj{\"{o}}strand}},
  \bibinfo{author}{\bibfnamefont{S.}~\bibnamefont{Mrenna}}, \bibnamefont{and}
  \bibinfo{author}{\bibfnamefont{P.}~\bibnamefont{Skands}},
  \bibinfo{journal}{J. High Energy Phys.} \textbf{\bibinfo{volume}{0605}},
  \bibinfo{pages}{026} (\bibinfo{year}{2006}).

\bibitem[{\citenamefont{{M.~Aliev, H.~Lacker, U.~Langenfeld, S.~Moch, P.~Uwer,
  M.~Wiedermann}}(2011)}]{Hathor}
\bibinfo{author}{\bibnamefont{{M.~Aliev, H.~Lacker, U.~Langenfeld, S.~Moch,
  P.~Uwer, M.~Wiedermann}}}, \bibinfo{journal}{{Comput.Phys.Commun.}}
  \textbf{\bibinfo{volume}{182}}, \bibinfo{pages}{1034} (\bibinfo{year}{2011}).

\end{thebibliography}

\clearpage 
\begin{flushleft}
{\Large The ATLAS Collaboration}

\bigskip

G.~Aad$^{\rm 48}$,
B.~Abbott$^{\rm 111}$,
J.~Abdallah$^{\rm 11}$,
A.A.~Abdelalim$^{\rm 49}$,
A.~Abdesselam$^{\rm 118}$,
O.~Abdinov$^{\rm 10}$,
B.~Abi$^{\rm 112}$,
M.~Abolins$^{\rm 88}$,
H.~Abramowicz$^{\rm 153}$,
H.~Abreu$^{\rm 115}$
% ...
\bigskip

% ...
$^{10}$ Institute of Physics, Azerbaijan Academy of Sciences, Baku, Azerbaijan\\
$^{11}$ Institut de F\'isica d'Altes Energies and Departament de F\'isica de la Universitat Aut\`onoma  de Barcelona and ICREA, Barcelona, Spain\\
% ...
$^{48}$ Fakult\"{a}t f\"{u}r Mathematik und Physik, Albert-Ludwigs-Universit\"{a}t, Freiburg i.Br., Germany\\
$^{49}$ Section de Physique, Universit\'e de Gen\`eve, Geneva, Switzerland\\
% ...
$^{88}$ Department of Physics and Astronomy, Michigan State University, East Lansing MI, United States of America\\
% ...
$^{111}$ Homer L. Dodge Department of Physics and Astronomy, University of Oklahoma, Norman OK, United States of America\\
$^{112}$ Department of Physics, Oklahoma State University, Stillwater OK, United States of America\\
% ...
$^{115}$ LAL, Univ. Paris-Sud and CNRS/IN2P3, Orsay, France\\
% ...
$^{118}$ Department of Physics, Oxford University, Oxford, United Kingdom\\
% ...
$^{153}$ Raymond and Beverly Sackler School of Physics and Astronomy, Tel Aviv University, Tel Aviv, Israel\\
% ...
\end{flushleft}

\end{document}